\documentclass[11pt]{article}
\pdfoutput=1 
\usepackage[margin=1in]{geometry}
\usepackage{graphicx}
\usepackage{booktabs}
\usepackage{amsmath}
\usepackage{amssymb}
\usepackage{array}
\usepackage{cmap}
\usepackage[T1]{fontenc}
\usepackage{helvet}
\input{glyphtounicode}\pdfgentounicode=1
\usepackage[expansion=false,protrusion=false]{microtype}
\DisableLigatures[f]{encoding = T1, family = *}
\usepackage{natbib}
\usepackage{xcolor}
\definecolor{linkblue}{HTML}{2A5DB0}
\definecolor{boxgray}{HTML}{F0F2F5}
\usepackage[colorlinks=true,citecolor=linkblue,linkcolor=linkblue,urlcolor=linkblue]{hyperref}

\makeatletter
\renewcommand\section{\@startsection{section}{1}{\z@}%
  {-3.5ex \@plus -1ex \@minus -.2ex}{2.0ex \@plus.2ex}%
  {\normalfont\large\sffamily\bfseries}}
\renewcommand\subsection{\@startsection{subsection}{2}{\z@}%
  {-2.5ex\@plus -1ex \@minus -.2ex}{1.2ex \@plus .2ex}%
  {\normalfont\normalsize\sffamily\bfseries}}
\renewcommand\paragraph{\@startsection{paragraph}{4}{\z@}%
  {1.6ex \@plus1ex \@minus.2ex}{-0.6em}%
  {\normalfont\normalsize\sffamily\bfseries}}
\makeatother
\newcommand{\abstractbox}[1]{%
  \begingroup\setlength{\fboxsep}{10pt}%
  \noindent\colorbox{boxgray}{\begin{minipage}{\dimexpr\textwidth-2\fboxsep\relax}#1\end{minipage}}%
  \endgroup}

\newcommand{\dash}{\textemdash}
\newcommand{\code}[1]{\texttt{#1}}

\begin{document}

\begin{flushleft}
{\sffamily\bfseries\fontsize{20}{24}\selectfont
Teach it to stop, not just to click\par}
\vspace{10pt}
{\sffamily\bfseries Barada Sahu}, Cabal AI \quad {\sffamily\bfseries Shivesh Pandey}, Para AI
\end{flushleft}

\abstractbox{%
Agentic computer-use RL is reported in single runs, and those numbers mislead. Using verifier-guided repair of a 35B computer-use agent (CUA) as a testbed across five oracle-graded environments, we show that a repaired policy's success rate is dominated by \emph{upstream} variance: a variance-components decomposition across three cells (crossed data-draw $\times$ seed grid, bootstrap CIs) finds evaluation variance negligible ($\sigma_{\mathrm{eval}}{\approx}0$) and the training-seed effect small everywhere ($\leq10\%$); instead it splits between the data draw and run-to-run nondeterminism, and the data draw's share grows with difficulty---a minority on the two reliably-repaired cells (where run-to-run variance is larger), rising to dominant ($48\%$, $\sigma_{\mathrm{data}}{=}0.26$) on the hardest. On that hardest cell the run-to-run distribution is bimodal (Hartigan dip $p{=}0.07$, $k{=}10$, one cell), so a single run has ${\sim}30\%$ chance of the failure mode and mean$\pm$std is the wrong summary. On that footing, two positive findings hold. First, repairability is two-tier in how \emph{constrained} the corrective action is: a single fixed token is installed reliably (done-detection $0.97{\pm}0.06$), while open-ended corrections are only partial---spatial-coordinate clicks (field-targeting $0.71{\pm}0.26$, grounding $0.53{\pm}0.35$) and, hardest of all, a generative field-fill ($0.14{\pm}0.04$). Second, the frame-level repair transfers to oracle-graded task success only when the corrective action is the task's sole remaining blocker (LinkedIn $8/20$ vs.\ base $0/15$, Fisher $p{=}0.006$). We caught two of our own over-claims---a sample-efficiency curve and a ``grounding cannot be bought'' boundary---only by replicating across seeds; and a stress test makes the stakes external: a single-run improvement of the size this field publishes would have the wrong sign ${\sim}1/3$ of the time \emph{if} run in a variance regime comparable to ours. We release a protocol and library (\code{cua\_reliability}) that make $k$-seed reporting routine. The apparatus behind the numbers is, to our knowledge, the first multimodal \emph{segment-aggregated on-policy self-distillation} (SA-OPSD) update on a real 35B CUA policy.

\vspace{4pt}
\textbf{Date:}~\today \\
\textbf{Correspondence:}~\texttt{barada@gmail.com, cs21bt067.alum25@iitdh.ac.in} \\
\textbf{Code:}~the SA-OPSD method, completion verifier, and the leakage-free held-out plus bootstrap evaluation harness are open-source; the oracle-graded environments and the 35B training/serving backend are proprietary and not released.
}
\vspace{6pt}

\section{Introduction}

The dominant reflex for improving a computer-use agent (CUA) is to \emph{tell it more}: add an instruction, refine a grounding module, drop an exemplar into the prompt. When that fails, practitioners escalate to fine-tuning or reinforcement learning, or wrap the policy in a corrective harness. These are five distinct \textbf{correction channels}---naive in-context prompting, verifier-guided in-context hints, imitation fine-tuning, reward-based updates, and inference-time override. Two of these live in the prompt or harness and must be re-supplied at every step (we call this \emph{renting} the fix); the other two write the fix into the weights (\emph{buying} it). This paper asks which channel repairs which failure, whether the repair reaches end-to-end task success, and how reliably each answer holds up under replication.

The apparatus tying the channels together is a \textbf{completion verifier}: a second, more capable VLM that judges---from the same observation the policy sees---whether the task is complete and what the corrective next action is. The verifier is the \emph{shared substrate} connecting channels: the same judgment can be an in-context hint, a distillation teacher, a reward, or an inference gate. With it we compare channels on equal footing across five environments.

Our central contribution is \textbf{reliability}. Agentic-CUA-RL results are reported in single runs, but those numbers are dominated by \emph{upstream} variance---in the data draw and run-to-run nondeterminism, not evaluation or the training seed, with the data draw's share rising to dominant on the hardest cell (\S\ref{sec:reliability}); we measure this, turn it into a $k$-seed protocol and a released library, and use it to catch two of our own over-claims. That lens is what lets the two positive findings stand. First, a \textbf{two-tier difficulty gradient}: verifier-guided weight updates repair every failure type above a near-zero base, but a single fixed token installs reliably while open-ended corrections (spatial-coordinate clicks, and a generative field-fill) are only partial. Second, the repair \textbf{transfers to end-to-end task success} exactly when the corrective action is the task's sole remaining blocker (LinkedIn $8/20$ vs.\ base $0/15$, Fisher $p{=}0.006$), and not otherwise. The verifier-guided repair (\S\ref{sec:method}) is the \emph{apparatus} that generates these numbers; we report mean and standard deviation over $k$ training seeds.

\paragraph{Contributions.}
\begin{enumerate}
\item A \textbf{measured reliability methodology for agentic CUA-RL} (\S\ref{sec:reliability})---the spine of the paper: a variance-components decomposition (evaluation ${\approx}0$, training seed $\leq10\%$, data draw dominant at $48\%$ on the hardest cell), a formal bimodality test, and a seed-budget power analysis, validated out-of-sample---plus two falsifications of our own claims. Released as \code{cua\_reliability}.
\item A \textbf{two-tier difficulty-gradient result} (\S\ref{sec:results}, Table~\ref{tab:robust}), on held-out independent trajectories over four cells: a fixed-token stop installs reliably (done-detection $0.97{\pm}0.06$) but open-ended corrections only partially and unreliably (field-targeting $0.71{\pm}0.26$, grounding $0.53{\pm}0.35$, generative field-fill $0.14{\pm}0.04$)---every rate with a bootstrap CI excluding its base; the near-vs-far coordinate gap seen on training frames collapses on held-out, and the token-vs-coordinate ordering replicates on a second, architecturally distinct policy (Fara-7B).
\item An \textbf{end-to-end transfer boundary} (\S\ref{sec:results}, Table~\ref{tab:e2e}), oracle-graded and warm-served: the frame-level repair completes tasks when its action is the sole blocker (LinkedIn $8/20$ vs base $0/15$, $p{=}0.006$), not otherwise. Getting a stable number required a load-once serving fix that removes the per-rollout reload.
\item The \textbf{apparatus} (\S\ref{sec:method}): the completion verifier as hint, teacher, reward, and gate; and the first \emph{multimodal SA-OPSD} update on a real 35B VLM policy---the tool that generates the repaired policies whose reliability we characterize.
\end{enumerate}

A prior version of this work claimed a narrow ``weights over words'' dissociation on one task, and an earlier draft of \emph{this} work claimed a hard ``rent-only'' boundary for spatial grounding. The replicated evidence here \textbf{retracts both strong claims} in favor of the graded, verifier-centric account.

\section{Setup}
\label{sec:setup}

\paragraph{Agent.} Holo3-35B-A3B, a vision-language MoE, run as a pure perception$\to$action loop (\code{/v1/cua}): the model receives the current screenshot plus a windowed textual action-history and emits one action per step (\code{click(x,y)}, \code{type\_text}, \code{scroll}, \code{key\_press}, \code{done()}). Only the current screenshot is sent (prior frames pruned to fit the 8192-token serving context); every model call---prompt, image, generated reasoning $+$ action, and per-token log-probabilities---is logged.

\paragraph{Environments and oracles.} Five deterministic, snapshot-resettable web-mirror environments (LinkedIn, Indeed, Fiverr, Mercor, Shopify), each with a fixed seed and frozen clock. \emph{Reset-per-rollout} gives every trial a byte-identical start. \emph{Programmatic grading}: each environment exposes an oracle (\code{/\_\_env\_\_/oracle?task\_id}) that reads database state and the audit log---never the transcript---plus a timestamped mutation log for segment-level ground truth. DB-level grading replaces the screenshot heuristics that make agent-RL results unreliable (\S\ref{sec:reliability}).

\paragraph{The completion verifier.} A separate VLM (Claude, \code{claude-sonnet-4-6}) is prompted as a task-completion judge: given the task and current screenshot, output \code{\{complete, evidence, next\}}. It is decoupled from the policy---it does not share Holo3's perceptual blind spots (\S\ref{sec:tax})---and is 3/3 on our completion probe, including cases the policy gets wrong. It plays four roles: in-context \emph{hint}, distillation \emph{teacher}, \emph{reward}, and inference \emph{gate}.

\paragraph{Metric.} Because §\ref{sec:reliability} shows single-run CUA numbers mislead, our headline results are \textbf{decision-state action-emission rates}: the fraction of the failure frames on which the tuned policy (adapter on) emits the corrective action, evaluated adapter-on vs.\ adapter-off on the same model load, on \emph{held-out independent trajectories}, and reported as \textbf{mean~$\pm$~std over five training seeds} with a bootstrap $95\%$ CI. Base-policy emission of each corrective action is at or near $0$ across all held-out frames ($0.00$ for done-detection and field-targeting; $0.05$ for spatial grounding).

\paragraph{On verifier circularity.} One completion verifier plays three roles that a skeptic will note are not independent: it \emph{selects} the decision frames (those where it recommends the corrective action), it \emph{supplies} the training label (teacher action), and its action is what the emission \emph{metric} counts. Measured this way, the frame-level rates answer a deliberately narrow question---can a weight update install the verifier's own judgment into the policy?---and nothing broader. We break the loop in two places, and lean on those for any claim that outruns it. First, the \textbf{end-to-end results (\S\ref{sec:results}, Table~\ref{tab:e2e}) are graded by the DB oracle}, which reads database state and the audit log and never consults the verifier; the LinkedIn transfer ($8/20$ vs $0/15$) is a verifier-independent number. Second, the verifier's judgment is \textbf{not idiosyncratic to one vendor}: on a completion probe four verifiers agree $\geq 14/16$ and an open 72B model ties closed GPT-4o with zero false-positives (Table~\ref{tab:verifiers}), so ``the corrective action'' is not a Claude-specific artifact. The frame-level rates are an instrumented proxy; the oracle-graded transfer is the load-bearing claim.

\section{Failure taxonomy}
\label{sec:tax}

We drive the base agent on the primary task of each environment and observe a mechanistically distinct failure in each (Table~\ref{tab:taxonomy}). Base task-success is near-zero across families (Table~\ref{tab:baserate}, six trials each); the one partial exception is Fiverr, where the agent usually completes the order ($4/6$) but intermittently keeps acting afterward---the over-execution that \emph{is} the done-detection failure, so even Fiverr's ``passes'' carry the target defect. Several failures are \emph{state-misperceptions}: the agent acts correctly given a false belief (LinkedIn ``already typed''; Fiverr ``not yet ordered''; Indeed ``remote already on''). This is why naive prompting is inert---a mistaken percept outranks an instruction.

\begin{table}[h]
\centering
\small
\begin{tabular}{@{}p{1.5cm}p{6.6cm}p{2.5cm}p{2.5cm}@{}}
\toprule
\textbf{Env} & \textbf{Base failure (from telemetry)} & \textbf{Class} & \textbf{Corrective action} \\
\midrule
LinkedIn & clicks \emph{Send} on empty note---``I already typed the note'' & action-selection & \code{type\_text} (token) \\
Shopify & email into \emph{Notes}, \emph{Merchant name} left empty $\to$ validation wall & field-targeting & \code{click} prominent field (near-prior) \\
Mercor & stuck on step 1 of 5; rate typed twice; fields empty & form-progression & \code{click}/\code{type} field \\
Fiverr & places correct order, \emph{then keeps going} $\to$ collateral order & done-detection & \code{done()} (token) \\
Indeed & never clicks \emph{Remote} chip; ``the Remote filter is enabled'' (false) & spatial grounding & \code{click(178,330)} (far-prior) \\
\bottomrule
\end{tabular}
\caption{Base-agent failures across five environments. The corrective action ranges from a discrete token to a near- or far-prior click coordinate.}
\label{tab:taxonomy}
\end{table}

\paragraph{Breadth of the base-failure regime.} The five failures above are not cherry-picked: across \emph{eight} oracle-graded app families---the five repair environments plus three further ones (Slack, Notion, HubSpot)---we ran the primary task six times each (base agent, reset-per-rollout), and the base success rate is $\approx 0$ almost everywhere (Table~\ref{tab:baserate}). Seven of eight apps sit at $0/6$; the lone exception is Fiverr done-detection at $4/6$---the agent usually completes the order but \emph{intermittently} over-executes, which is exactly why done-detection is a repair target rather than a solved case. HubSpot is excluded: all six rollouts fast-failed at $\sim$45\,s (infrastructure, not agent behavior), leaving no decision trajectory. Because base success is $\approx 0$ almost everywhere, every non-zero repaired rate below is a real effect, not regression to a nonzero prior.

\begin{table}[h]
\centering
\small
\begin{tabular}{@{}p{1.8cm}p{4.4cm}cp{3.2cm}@{}}
\toprule
\textbf{App} & \textbf{Primary task} & \textbf{Base pass} & \textbf{Failure class} \\
\midrule
LinkedIn & connect / note & $0/6$ & action-selection \\
Indeed & search $+$ save w/ Remote & $0/6$ & spatial grounding \\
Fiverr & order Basic logo, stop & $4/6$ & done-detection (intermittent) \\
Mercor & apply to ML engineer & $0/6$ & form-progression \\
Shopify & submit Plus lead & $0/6$ & field-targeting \\
Slack & post message & $0/6$ & action-selection \\
Notion & create page & $0/6$ & form-progression \\
HubSpot & create contact & \dash & infra fast-fail (excluded) \\
\bottomrule
\end{tabular}
\caption{Base success across eight app families, six reset-per-rollout trials each, oracle-graded. Seven of eight sit at $0/6$; Fiverr is the sole partial ($4/6$, intermittent over-execution). HubSpot's six trials all fast-failed at $\sim$45\,s (infrastructure) and are excluded. The near-universal zero base rate is what the repair rates in \S\ref{sec:results} are measured against.}
\label{tab:baserate}
\end{table}

\section{Apparatus: the verifier as a shared substrate, and multimodal SA-OPSD}
\label{sec:method}

The repair method below is the \emph{apparatus} of this paper---the tool that produces the repaired policies whose reliability \S\ref{sec:reliability} characterizes. We describe it in full because the reliability results only mean something alongside the exact update that produced them.

\paragraph{Verifier-as-hint (smart rent).} In a harness loop, after each step we inject the verifier's \code{next} recommendation into the observation; the policy re-generates.

\paragraph{Verifier-as-reward-and-teacher (SA-OPSD).} Our weight update is \emph{segment-aggregated on-policy self-distillation} (SA-OPSD), in the on-policy-distillation family~\citep{gkd,vlaopd}: for a group of on-policy steps sharing a \code{group\_id}---the \emph{segment} that the ``SA'' names---the verifier scores each step (acting after completion $=$ over-execution $=$ reward 0; progress $=$ reward 1), yielding group-relative advantages $A_i$ aggregated within the segment. The loss combines two terms:
\begin{equation}
\mathcal{L} = \underbrace{\mathrm{GRPO}\!\left(A_i,\ \log\pi_\theta,\ \log\pi_{\mathrm{old}}\right)}_{\text{down-weight bad actions}} \;+\; \underbrace{\sigma\!\left(-A_i/\tau\right)\cdot \mathrm{BC}\!\left(a^{\mathrm{teacher}}\right)}_{\text{install the verifier's corrected action}} .
\end{equation}
The GRPO term uses the clipped importance ratio $\rho_t=\exp(\log\pi_\theta-\log\pi_{\mathrm{old}})$ against the captured $\log\pi_{\mathrm{old}}$, with a Schulman-$k_3$ KL anchor to the base (LoRA-adapter-disabled). The distillation term is behavior cloning of the verifier's corrected action, gated to fire where the student is failing. \textbf{The policy log-probabilities are conditioned on the screenshots}---the multimodal forward is what lets this touch a VLM policy rather than a text LM. The reward channel supplies a \emph{negative} on bad actions, which imitation cannot express; the teacher supplies a \emph{positive} the base never explores. These are complementary, and which one carries a repair depends on the failure: for done-detection, where the base never samples \code{done()}, the distillation term is what installs it and the reward term alone cannot---we isolate the two terms in the same trainer and confirm this (Table~\ref{tab:attribution}).

\paragraph{Verifier-as-gate (deployment).} At inference, a policy proposal of \code{done()} (or any irreversible action) is accepted only if the verifier confirms completion---a cheap safety valve against over-correction (\S\ref{sec:results}).

\paragraph{Verifier-as-override.} For the hardest cases, the harness may execute the verifier's corrective coordinate directly (the \code{ground\_clicks}/director pattern), bypassing the policy---now a deployment convenience rather than a necessity (\S\ref{sec:results}).

\section{Results: a difficulty gradient}
\label{sec:results}

\paragraph{Naive rent fails universally; smart rent splits by failure.} A static instruction---even a literal in-prompt exemplar---produces $0$ correct emissions in $\sim$80 modal iterations on the LinkedIn note. Injecting the verifier's hint in-context \emph{flips} the Fiverr done-detection misperception (base: ``the button isn't visible$\dots$ scroll''; hinted: ``$\dots$confirming the order was successfully placed'') but is \emph{ignored} on Indeed grounding, where the false percept (``the Remote filter is enabled'') overrides the advice even with explicit coordinates. So verifier-guided \emph{rent} repairs semantic failures in-context but not perceptual-grounding ones.

\paragraph{Held-out measurement (leakage-free).} \S\ref{sec:reliability} shows that scoring on the \emph{same} trajectory's frames that a checkpoint trained on inflates and destabilizes the number (pseudo-replication: frames from one rollout are correlated). Our headline numbers therefore come from an \textbf{independent held-out set}: for each cell we run \emph{fresh} base rollouts, exclude the trajectory used for training (leakage-free run-level split), and keep only the \emph{decision frames}---those where the completion verifier recommends the corrective action. Each of the five seed checkpoints is evaluated on this held-out set (Indeed $21$ decision frames / $6$ trajectories; Fiverr $34$/$8$; Shopify $9$/$9$), and we report mean~$\pm$~std over seeds plus a bootstrap $95\%$ CI over the pooled (seed~$\times$~frame) hits. Because the held-out set is decision-frames-only (not diluted by easy progress frames), these rates are both more valid \emph{and} higher than the training-frame rates of an earlier draft.

\paragraph{The main result: weight updates repair every failure, at graded difficulty.} Table~\ref{tab:robust} reports the held-out corrective-action emission rate, mean~$\pm$~std over five training seeds with a bootstrap $95\%$ CI, against the base rate on the same held-out frames.

\begin{table}[h]
\centering
\small
\begin{tabular}{@{}p{4.6cm}p{1.4cm}cccc@{}}
\toprule
\textbf{Failure (corrective action)} & \textbf{Channel} & \textbf{Held-out} & \textbf{Boot} & \textbf{Base} & \textbf{$n$} \\
& & \textbf{rate} & \textbf{95\% CI} & & \\
\midrule
Done-detection (Fiverr \code{done()}) & SA-OPSD & $\mathbf{0.97 \pm 0.06}$ & $[0.95,\,0.99]$ & $0.00$ & $170$ \\
Field-targeting (Shopify Merchant, \emph{near}-prior click) & SFT & $\mathbf{0.71 \pm 0.26}$ & $[0.58,\,0.84]$ & $0.00$ & $45$ \\
Spatial grounding (Indeed Remote chip, \emph{far}-prior click) & SA-OPSD & $\mathbf{0.53 \pm 0.35}$ & $[0.44,\,0.63]$ & $0.05$ & $105$ \\
Form-progression (Mercor Headline, \emph{generative} field-fill) & SFT & $\mathbf{0.14 \pm 0.04}$ & $[0.08,\,0.19]$ & $0.00$ & $140$ \\
\bottomrule
\end{tabular}
\caption{The difficulty gradient, measured on \emph{held-out independent trajectories} (decision frames only, \textbf{five} training seeds, bootstrap $95\%$ CI over pooled seed$\times$frame hits, $n$ = pooled decision frames). The axis is how \textbf{constrained} the corrective action is: a single \emph{fixed} token (\code{done()}) is installed reliably ($0.97$, CI $[0.95,0.99]$); \emph{open-ended} corrections are only partial---the two spatial-coordinate clicks are high-variance ($0.53$--$0.71$, $\pm 0.26$--$0.35$, one grounding seed at $0.00$; the near-vs-far distinction that training frames suggested \emph{collapses} on held-out data), and a \emph{generative} field-fill (locate the empty field \emph{and} produce apt content) is the hardest cell of all ($0.14$). All four CIs exclude their base rate.}
\label{tab:robust}
\end{table}

\begin{figure}[h]
\centering
\includegraphics[width=0.82\linewidth]{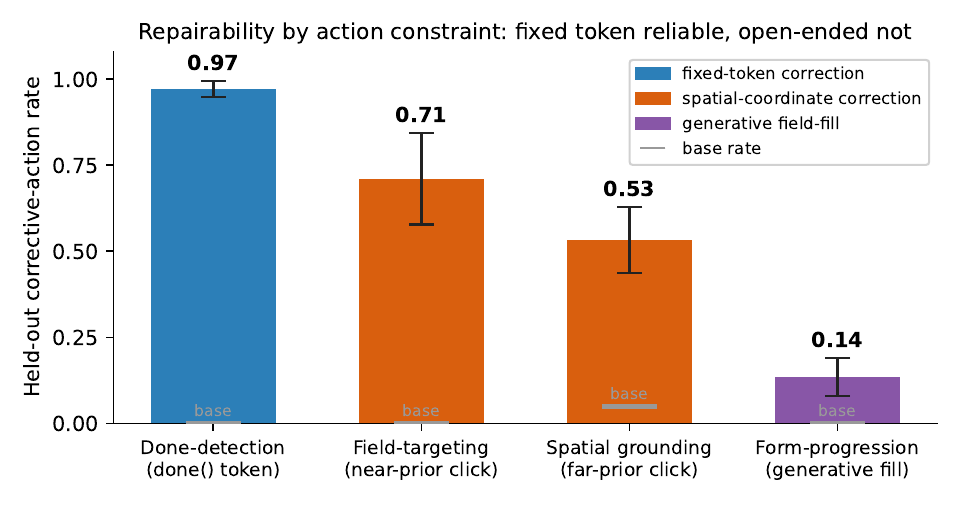}
\caption{Held-out corrective-action rate (five seeds, bootstrap $95\%$ CI) against base. A fixed-token correction is installed reliably; open-ended corrections sit far lower---both spatial-coordinate clicks (near- or far-prior) with wide CIs, and the generative field-fill (Mercor) lowest of all.}
\label{fig:gradient}
\end{figure}

The ordering tracks the corrective-action type. A discrete-token stop (Fiverr \code{done()}, $0.97 \pm 0.06$) has a tight CI, $[0.95,0.99]$. The two spatial-coordinate corrections are both lower and unstable: field-targeting (Shopify, $0.71 \pm 0.26$; runs $0.22/1.00/0.78/0.78/0.78$) and grounding (Indeed, $0.53 \pm 0.35$; runs $0.81/0.24/0.86/\mathbf{0.00}/0.76$), each with a seed spread of $\pm 0.26$--$0.35$. Training-frame numbers had suggested a clean near-vs-far gap ($0.92$ vs $0.36$); on independent trajectories that gap disappears, and coordinate generation is set by seed-to-seed training variance rather than by distance from the visual prior. Below the coordinates sits the Mercor \emph{generative field-fill} ($0.14 \pm 0.04$, base $0.00$): installing a \code{type\_text} that must both target the empty Headline and generate apt content is the hardest correction we test. The axis is thus not token-vs-coordinate per se but how \textbf{constrained} the corrective action is: \textbf{a single fixed token is reliably buyable; open-ended corrections---spatial coordinates, and generative fills most of all---are only partially and unreliably buyable}. All four held-out CIs exclude the base rate, so each repair is resolved rather than one lucky run. (The LinkedIn action-selection repair, from prior work, is $3/4$ end-to-end against the DB oracle; as a single-run number it carries the caveat of \S\ref{sec:reliability}.)

\paragraph{What sets the gradient---and what does not.} A natural hypothesis is that repairability tracks the base model's \emph{prior} on the corrective action: the more probability the untuned policy already places on the fix, the easier it is to install. We measured it---the base model's mean per-token log-prob of the corrective action at each decision frame---and the relationship is inverted: the most reliably-installed correction, \code{done()} at $0.97$, has by far the \emph{lowest} base prior ($-13.6$ vs.\ $-3.2$ and $-3.4$ for the two coordinate clicks), because at the over-execution frame the base is committed to clicking and assigns stopping almost no mass. So repairability is not governed by how much the base already ``wants'' the corrective action; it is governed by the target's \textbf{discreteness}. A single fixed token installs reliably \emph{even against a strong prior toward acting}, whereas a precise coordinate resists reliable install \emph{even though the base emits clicks constantly}---the difficulty is the continuous, precise target, not the base's willingness. (The prior is a whole-string mean, so it also reflects action-format frequency: the base's click tokens are cheap---yet the token cell, hardest by prior, is the easiest to repair.)

\paragraph{The gradient replicates on a second policy.} To test whether the two-tier structure is specific to Holo3's tokenizer and action head, we repeat two cells on \textbf{Fara-7B}, an architecturally distinct CUA---a dense $7$B Qwen2.5-VL, versus Holo3's $35$B mixture-of-experts. We roll Fara out on its own trajectories, build a leakage-free held-out from its decision frames, and SFT it on the corrective action. From a $0.00$ base on both cells, the fixed-token stop installs at $1.00 \pm 0.00$ over three seeds while the coordinate click installs at $0.81 \pm 0.14$: the same ordering, and the same variance signature (the token cell tight, the coordinate cell not). The structure is not a Holo3 artifact. Magnitudes are policy-specific---Fara installs the coordinate more reliably than Holo3 ($0.81$ vs $0.53$)---so what generalizes is the constrained-vs-open-ended \emph{ordering}, not the exact rates. Held-out $n$ is small ($11$ and $7$ frames), so this is directional; the point is the replicated structure.

\paragraph{A note on held-out sample sizes.} The cells contribute different $n$ because the failure types recur at different rates within a trajectory: done-detection and grounding decision frames recur across many steps (Fiverr $n{=}170$, Indeed $n{=}105$ pooled over five seeds), and the Mercor fill recurs while the base stays stuck ($28$ frames from $10$ trajectories, pooled $n{=}140$), whereas a field-targeting correction fires \emph{once} per trajectory (the field is filled and then left alone), so Shopify's leakage-free held-out set is necessarily small---$9$ frames, one per independent trajectory, pooled to $n{=}45$ over five seeds. Its wider CI reflects that scarcity, not a weaker effect.

\paragraph{The Mercor cell, and one more inflation.} Mercor's form-progression repair (Table~\ref{tab:robust}, bottom row) is the fourth cell, on a freshly built leakage-free held-out ($28$ decision frames from $10$ independent base trajectories). It is real but weak: base fills the Headline in $0/28$ frames, the $k{=}5$ SFT adapter in $0.14 \pm 0.04$ (every seed in $[0.07,0.18]$, CI $[0.08,0.19]$ excluding base). It also furnishes a third in-domain instance of the inflation this paper warns about: a single-seed number scored on \emph{training} frames read $0.40$, nearly $3\times$ the leakage-free $k{=}5$ value---the same training-frame optimism as the near-vs-far coordinate gap.

\paragraph{Which term installs the stop? The distillation term.} SA-OPSD carries two terms---a GRPO policy term and advantage-gated distillation of the verifier's \code{done()}---and it is natural to ask which one does the work. We isolate each in the \emph{same} trainer on the same eight-frame Fiverr batch, matched on gating and epochs (the switch zeroes one term's contribution), and score \code{done()}-emission on the held-out decision frames (Table~\ref{tab:attribution}). \textbf{The reward term alone} (GRPO-alone, teacher stripped) installs nothing---$0.00 \pm 0.00$ over three seeds ($n{=}34$ frames each); it moves the policy off the harmful second click toward \code{scroll} and re-clicks but never \code{done()}, because \code{done()} appears in $0/15$ base frames and an importance-ratio reward term can only up-weight actions the policy already samples---it has no \code{done()} mass to reinforce. \textbf{The distillation term alone} (GRPO term switched off, gated BC of the teacher \code{done()}) installs the stop completely---$1.00 \pm 0.00$ over three seeds---the teacher supplies the positive the base never explores, and behavior-cloning it generalizes to the held-out frames. Adding the reward term back (full SA-OPSD, $0.97$) does \emph{not} improve on distillation alone. So on done-detection the \textbf{distillation term is the active ingredient}: the reward term's role is elsewhere---down-weighting harmful actions the policy \emph{does} sample (the collateral clicks the coordinate cells must suppress), which imitation cannot express. One caveat frames the number: the held-out is decision-frames-only (all are over-execution points where \code{done()} is correct), so this rate measures whether the action is \emph{installed}, not whether its \emph{timing} is correct---distillation's tendency to over-emit \code{done()} is exactly what the verifier gate handles (\S\ref{sec:results}, ``Over-correction and the gate'').

\begin{table}[h]
\centering
\small
\begin{tabular}{@{}p{6.4cm}cc@{}}
\toprule
\textbf{Condition (done-detection)} & \textbf{\code{done()}-rate} & \textbf{seeds} \\
\midrule
Base (no repair) & $0.00$ & --- \\
Reward term only (GRPO-alone) & $0.00 \pm 0.00$ & $3$ \\
\textbf{Distillation term only} (matched, gated) & $\mathbf{1.00 \pm 0.00}$ & $3$ \\
SA-OPSD (reward $+$ gated distill) & $0.97 \pm 0.06$ & $5$ \\
\bottomrule
\end{tabular}
\caption{Term-attribution on Fiverr done-detection (held-out \code{done()}-emission on decision frames, $n{=}34$/seed; each term isolated in the \emph{same} trainer with the other switched off, matched batch/gating/epochs). The reward term alone cannot install a token the base never samples ($0.00$); the distillation term alone installs it ($1.00$); combining them does not beat distillation alone. The rate measures \emph{installation}, not timing---all frames are over-execution points where \code{done()} is correct, and distillation's over-emission is what the gate corrects.}
\label{tab:attribution}
\end{table}

\paragraph{Over-correction and the gate.} At small data (8 examples) SA-OPSD over-generalizes \code{done()} to some incomplete frames (premature stop). The \emph{weights$+$gate} combination resolves this: weights make \code{done()} available; the verifier gate (3/3 completion judgment) accepts it only when the task is complete---correct done-\emph{timing} that neither weights-alone nor base achieves.

\paragraph{End-to-end task success, and when the repair transfers.} Decision-frame emission is a proxy; the deployment question is whether a repaired policy completes more tasks. We serve each repaired policy (merged GGUF) and the base against the DB oracle through the pure-CUA loop, $5$ rollouts per arm, both arms warm-served. (Warm serving matters: the production server reloads the served model per rollout, and that cold-load dominates the signal---identical configs gave challenger $5/5$ then $0/5$. A load-once server, reused across rollouts, removes it; the numbers below are stable across repeats.) The result, Table~\ref{tab:e2e}: the repair transfers to task success \emph{when the repaired action is the task's sole remaining blocker}. On LinkedIn the corrective action is the note text, and base never completes the task ($0/15$: it sends an empty note or fails to connect); the repaired policy completes the connection-with-note $8/20$ ($40\%$, $95\%$ CI $[0.22,0.61]$; Fisher exact $p{=}0.006$ against base). It does not transfer where the repaired action is not the bottleneck (Fiverr: base never over-executes here, so done-detection has nothing to fix; all base failures are incomplete checkouts) or is one of several required steps (Shopify: filling one field does not complete a multi-field submission). The rollout counts are asymmetric by design, not by selection: once LinkedIn showed a non-null effect we spent additional rollouts there ($20$ challenger / $15$ base) to tighten the interval and reach significance, whereas the two null cells were already categorical at $5$ per arm (Fiverr $2/5{=}2/5$, Shopify $0/5{=}0/5$)---more trials on a $0$-vs-$0$ tie would not change the conclusion.

\begin{table}[h]
\centering
\small
\begin{tabular}{@{}p{5.8cm}ccc@{}}
\toprule
\textbf{Cell (repaired action)} & \textbf{Challenger} & \textbf{Base} & \textbf{Transfers?} \\
\midrule
LinkedIn (type note --- \emph{sole} blocker) & $\mathbf{8/20}$ & $0/15$ & yes ($p{=}0.006$) \\
Fiverr (\code{done()} --- base never over-executes) & $2/5$ & $2/5$ & no \\
Shopify (field-fill --- 1 of several steps) & $0/5$ & $0/5$ & no \\
\bottomrule
\end{tabular}
\caption{End-to-end oracle-graded task success, warm-served. The frame-level repair transfers to task completion only when the corrective action is the sole remaining blocker: on LinkedIn the repaired policy completes the task $8/20$ ($40\%$) while base never does ($0/15$; Fisher exact $p{=}0.006$). Where the action is not the bottleneck (Fiverr) or one of several required steps (Shopify), it does not transfer ($5$ rollouts/arm there).}
\label{tab:e2e}
\end{table}

\section{Reliability: two falsifications}
\label{sec:reliability}

Agent-RL numbers are easy to over-read. We report two of our own over-claims; they make the case for replication.

\paragraph{Measurement fidelity.} Early pilots scored success from screenshots, which was too ambiguous to use; oracle and mutation-log grading gave a clean signal.

\paragraph{Falsification 1: a curve.} A clean single-seed sample-efficiency curve (held-out type-rate $0.47{\to}0.82$ across $n{=}4{\to}16$) \emph{did not survive replication}: three data-seeds give non-monotonic means $0.44{\pm}0.16$, $0.25{\pm}0.15$, $0.40{\pm}0.24$ at $n{=}4,16,48$. The decisive control---\emph{same trainer, data, config, run twice}---gave $0.18$ vs.\ $0.68$ (a $0.5$ swing), while re-evaluating a fixed checkpoint is deterministic ($0.176$ both times). So the noise is \textbf{upstream of evaluation}---in training run-to-run nondeterminism and, as the variance-components decomposition below shows, even more in \emph{which data} one trains on.

\begin{figure}[h]
\centering
\includegraphics[width=0.82\linewidth]{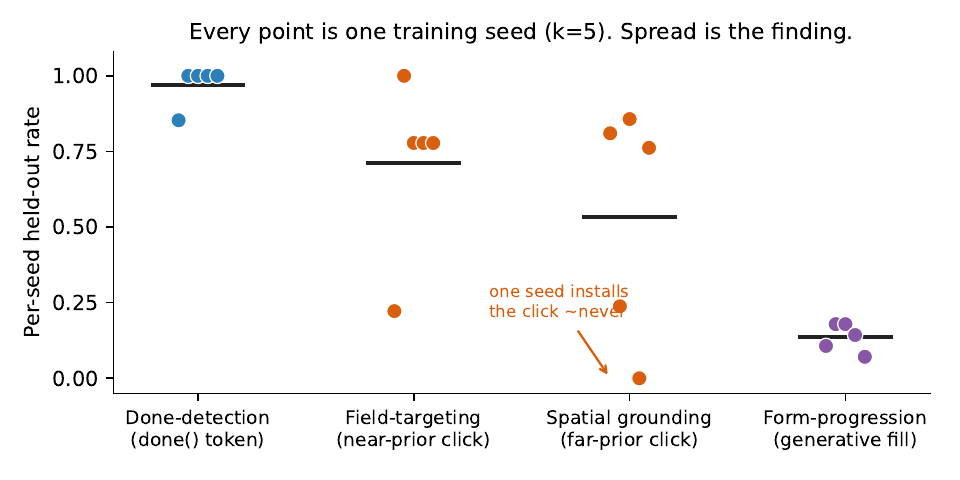}
\caption{Per-seed held-out rate ($k{=}5$); the bar is the mean. The discrete-token repair is stable across seeds; the spatial-coordinate repairs are not, and one grounding seed lands essentially at zero---the low tail that produced the earlier ``0/14'' boundary claim.}
\label{fig:seedvar}
\end{figure}

\paragraph{Falsification 2: a boundary.} An earlier draft of this paper reported the Indeed spatial-grounding repair as \textbf{0/14} across four weight-based attempts and concluded it \emph{cannot be bought}---a hard rent-only boundary. Under five-seed replication that same SA-OPSD update emits the Remote-chip click at $\mathbf{0.36 \pm 0.10}$ on the training frames, and at $\mathbf{0.53 \pm 0.35}$ (runs $0.81/0.24/0.86/0.00/0.76$; base $0.05$; bootstrap CI $[0.44,0.63]$) on the leakage-free held-out trajectories. \textbf{The 0/14 was one draw of a high-variance quantity}: across five seeds the same update ranges from $0.00$ to $0.86$, wider than the earlier draft's entire claimed effect. That one seed lands near $0/14$ is not a boundary but the low tail of a distribution whose mean sits well above base. Our most distinctive claim did not survive replication; by implication the earlier single-run negatives (SFT $0/5$, vision-LoRA $0/5$) are underpowered draws, not zeros. Figure~\ref{fig:seedvar} plots every seed: done-detection clusters tightly near $1$, while grounding scatters from $0.00$ to $0.86$. A single run could have sampled any point on that spread.

\paragraph{The protocol, validated out-of-sample.} Oracle grading; deterministic resets; leakage-free run-level held-out splits; and \textbf{mean~$\pm$~std over $k$ training runs per cell}. The one invariant across 12+ trainings is that base emission is $0$; the \emph{existence} of each weight effect is robust, its \emph{magnitude} is what requires $k$ runs. We validate the recommendation empirically rather than only deriving it: splitting each cell's seeds into independent halves, a $k$-seed mean predicts the held-out half's mean $1.3$--$1.7\times$ more accurately than a single run, which misestimates the true mean by up to $0.34$--$0.42$ ($90$th percentile) on the token and far-coordinate cells. Reporting $k$ seeds changes the answer, not just the error bars.

\paragraph{What this implies for the field's single-run deltas.} Agentic-RL improvements are routinely reported from a single run or a handful---process-reward and milestone methods claim gains from a few points to ${\sim}0.2$ in task success (e.g.\ $+7.7$ on AndroidWorld~\citep{guishepherd}). We cannot re-run those experiments, but we can stress-test the \emph{practice}: combine an improvement of the magnitude they report with the run-to-run variance we \emph{measure} in this domain, and ask how often a single-run measurement of it would have the wrong sign. In the high-variance coordinate regime ($\sigma{\approx}0.3$), a $+7.7$-point improvement is reported with the wrong sign $33$--$44\%$ of the time (empirical bootstrap on our measured bimodal distribution; Gaussian model), and even a large $+0.2$ gain flips $25$--$34\%$. A single-run delta of the size the field publishes is, in this regime, close to a coin flip on its sign---the external case for the protocol: without $k$-seed reporting, an improvement measured in a comparable variance regime could reverse on replication.

\paragraph{A measured variance budget.} We decompose the spread into its sources on all three cells with a crossed grid---$D{=}3$ \emph{independent} data draws (fresh disjoint base-rollout batches) $\times\,k{=}8$ training seeds, each checkpoint re-evaluated---and fit the variance components (Henderson crossed-ANOVA, parametric-bootstrap CIs; Table~\ref{tab:vc}). Two invariants hold across every cell: \textbf{$\sigma_{\mathrm{eval}}\approx 0$}, so the noise is never in evaluation (re-evaluating a fixed checkpoint is deterministic---confirmed on all re-eval pairs), and \textbf{$\sigma_{\mathrm{train}}$ is small} ($0.07$--$0.12$, at most $10\%$ of variance), so reseeding is never the main source---the naive ``training-run nondeterminism dominates'' story is wrong in every cell. The variance lives \emph{upstream}, in the data draw and the run-to-run residual. Data composition's share grows with difficulty: $\sim$$16$--$22\%$ on the two reliably-repaired cells, but $48\%$ (dominant) on the hardest, where $\sigma_{\mathrm{data}}{=}0.26$ exceeds every other component. \emph{Which} fresh batch you train on---chiefly how much corrective signal it happens to contain---matters most exactly where the repair is marginal. (This is also why the grid's far-coordinate mean, $0.29$, sits below the $0.53$ of the five-seed cell in Table~\ref{tab:robust}: the two use different independent data draws, and that gap \emph{is} $\sigma_{\mathrm{data}}$ in action.) A methodological corollary we verified in passing: measuring $\sigma$ on \emph{subsamples} of one batch rather than independent draws halves it ($0.18$ vs $0.36$ on the hard cell) and erases the catastrophic-failure mode entirely ($0/24$ vs $3/10$ runs below $0.10$)---how you construct the ``data seed'' can hide the very instability you are trying to measure.

\begin{table}[h]
\centering
\small
\begin{tabular}{@{}lccccc@{}}
\toprule
\textbf{Cell} & \textbf{$\sigma_{\mathrm{data}}$} & \textbf{$\sigma_{\mathrm{resid}}$} & \textbf{$\sigma_{\mathrm{train}}$} & \textbf{$\sigma_{\mathrm{frame}}$} & \textbf{$\sigma_{\mathrm{eval}}$} \\
\midrule
Fiverr (token, mean $0.79$)      & $0.10$ & $0.18$ & $0.07$ & $0.12$ & $\approx0$ \\
Shopify (near-coord, mean $0.72$) & $0.10$ & $0.13$ & $0.065$ & $0.13$ & $\approx0$ \\
Indeed (far-coord, mean $0.29$)   & $\mathbf{0.26}$ & $0.22$ & $0.12$ & $0.10$ & $\approx0$ \\
\bottomrule
\end{tabular}
\caption{Variance-components decomposition across three cells ($D{=}3$ independent data draws $\times$ $k{=}8$ seeds, Henderson crossed-ANOVA, bootstrap CIs). Two invariants: evaluation variance is a measured zero and the training-seed effect is small ($\leq10\%$) everywhere. Data composition ($\sigma_{\mathrm{data}}$) grows with cell difficulty---moderate on the reliably-repaired cells, dominant ($48\%$) on the hardest. To stabilize an agentic-RL number, control the data draw first, the seed second, and average out the residual.}
\label{tab:vc}
\end{table}

\begin{figure}[h]
\centering
\includegraphics[width=0.80\linewidth]{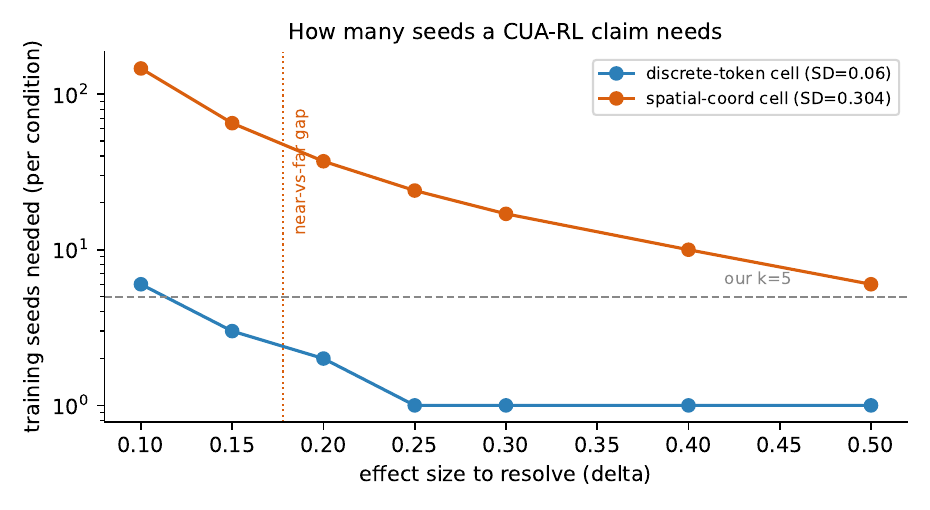}
\caption{Training seeds needed (per condition) to resolve an effect of size $\Delta$ at $80\%$ power, from the measured per-cell $\sigma$. Existence claims sit below our $k{=}5$; resolving the near-vs-far coordinate gap ($\Delta{=}0.18$) needs $\sim$$46$.}
\label{fig:power}
\end{figure}

\paragraph{Bimodality, and the seeds a claim needs.} The run-to-run distribution on the hardest cell is not Gaussian but \textbf{bimodal}---a Hartigan dip test rejects unimodality ($p{=}0.07$ on $10$ seeds; a two-component mixture is preferred by $\Delta\mathrm{BIC}{=}13$), with modes at $\approx0.09$ ($40\%$ of runs) and $\approx0.80$ ($60\%$). So mean$\pm$std is the wrong summary; we report the interquartile mean with a stratified bootstrap ($\mathrm{IQM}{=}0.56$, $95\%$ CI $[0.19,0.81]$) in the rliable style~\citep{rliable}. A single run has $\sim$$30\%$ chance of the failure mode---the mode the earlier ``$0/14$'' draw came from. This is a property of \emph{training runs}, not of which frames we score: restricting the held-out pool to the $16$ frames where the verifier specifically recommends the remote-chip click (vs.\ $5$ typing/other frames), the install-rate still ranges $0.0$--$0.94$ across fits with $7/23$ catastrophic---so the ``boundary'' is the low tail of a run-to-run distribution, not a wall the repair cannot cross and not an artifact of pooling. Bimodality is itself a seed-budget argument: at $k{=}5$ the dip test is underpowered ($p{=}0.38$); it takes the tenth seed to make the second mode visible. A power calculation then splits claims in two (Figure~\ref{fig:power}): \emph{existence} (a repair beats base) is cheap ($k{\leq}5$), but \emph{resolving} two high-variance cells apart ($\Delta{\approx}0.18$, $\sigma{\approx}0.3$) needs $k{\approx}46$---which is why we report the two coordinate cells as one tier. The rule, released as \code{cua\_reliability}: set $k$ from the cell's $\sigma$ and the effect claimed.

\paragraph{Is the grounding bimodality a representational limit or a training-reliability one?} We test the coordinate-tier resistance directly, and it is a \emph{reliability} phenomenon, not a wall. \emph{Exposure bias:} sampling $20$ generations per frame at temperature $1.0$, a trained (good-seed) adapter emits the correct Remote-chip coordinate on $100\%$ of held-out frames ($78\%$ of samples), and even the untuned base emits it on $14\%$ (though only $1.4\%$ of samples)---the coordinate is a very-low-probability action, not one the policy cannot represent. \emph{Capacity:} neither LoRA rank $\in\{8,16,32,64\}$ nor training-set size $n\in\{7,10,14\}$ systematically raises the install rate---both stay dominated by run-to-run instability (rank $8$ spans $0.57/0.00$; rank $64$, $0.62/0.10$; and the \emph{identical} rank-16 config, retrained, gave $0.05$ then $0.86$). So a good run at any capacity installs the coordinate, and nothing we vary controls whether a run is good. The mechanism ties to the base prior (\S\ref{sec:results}): because Holo3 places almost no mass on the coordinate, training must build that mass from near-nothing, which succeeds on some runs and bimodally collapses on others. \emph{And the instability is policy-specific, not a coordinate property}: the second policy, Fara-7B---a click-heavy CUA---installs the same coordinate \textbf{reliably} ($0.71$--$1.00$ across the four ranks, with a clean capacity effect and no collapse). So ``grounding resists weights'' is precisely \emph{grounding resists \textbf{reliable} weight-installation on a policy whose base prior for the target is near zero}---not a representational wall, and not a universal property of spatial-coordinate corrections. (This is one grounding cell across two policies and two capacity axes; we probed three further tasks for a clean second far-target grounding decision and found none---one was action-selection, one multi-target navigation, one infrastructurally unavailable---so a clean multi-cell version needs \emph{purpose-built} grounding tasks rather than harvested ones, which we leave to future work.)

\section{Related work}

\paragraph{Verifiers for computer-use agents.} A recent line of work builds separate verifiers that judge whether a GUI task succeeded, moving beyond passive screenshot checks. Agentic Reward Modeling lets a verifier agent run read-only probes and raises OSWorld-Verified accuracy from $84.7$ to $92.9$~\citep{agenticrm}; \citet{cuaverifiers} study verifier construction for computer-use agents directly. These use the verifier for \emph{evaluation} or as a reward. We use one completion verifier as a \emph{shared substrate} across four correction channels at once---in-context hint, distillation teacher, reward, and inference gate---which, to our knowledge, has not been done for CUAs.

\paragraph{Process and milestone rewards.} Rather than reward only the final outcome, dense reward models score each step: GUI-Shepherd trains a process reward model that adds $+7.7$ success on AndroidWorld~\citep{guishepherd}, and Adaptive Milestone Reward roughly doubles success on hard tasks~\citep{admire}. Multi-agent critics split evaluation into rule-based and semantic checks~\citep{osthemis,magicgui}. Our verifier supplies a lighter, single-model signal: a binary completion judgment that yields group-relative advantages and a distillation target, without a trained reward model or a critic ensemble.

\paragraph{On-policy distillation for VLM/VLA policies.} Teacher-student distillation during live interaction reduces exposure bias: VLA-OPD bridges offline SFT and online RL for vision-language-action models~\citep{vlaopd}, and Refined Policy Distillation turns generalist policies into RL experts~\citep{refinedpd}. SA-OPSD is in this family (GKD~\citep{gkd}; GRPO group-relative credit~\citep{grpo}) but pairs a GRPO policy term with advantage-gated distillation of a \emph{verifier's} corrected action, and is, to our knowledge, the first such multimodal update run on a real 35B computer-use policy.

\paragraph{Verifiable synthetic environments.} Code-native reward injection turns GUI traces into fast, deterministic environments whose reward is read from program state~\citep{guigenesis}. Our oracle-graded web mirrors are in the same spirit---each environment grades from database state and an audit log, never the transcript---which is what makes the reliability analysis below possible.

\paragraph{Reproducibility in RL.} \citet{deeprl} showed deep-RL results are fragile to seeds and implementation. We extend that lesson to agentic CUA settings with three concrete in-domain falsifications of our own claims, and argue $k$-run reporting is not optional here.

Our distinguishing contribution is the first in-domain \emph{measured} reliability characterization for agentic CUA-RL---a variance-components decomposition, a formal bimodality test, and a $k$-seed protocol---on a repair testbed, yielding a replicated difficulty-gradient and transfer result rather than single-run deltas.

\section{Limitations}
\begin{itemize}
\item \textbf{Two policies, not many.} The variance decomposition and the end-to-end results are on one agent, Holo3-35B-A3B; we replicate only the two-tier \emph{gradient} on a second, architecturally distinct policy (Fara-7B, a dense Qwen2.5-VL; \S\ref{sec:results}), where the ordering and its variance signature hold but the absolute magnitudes differ. So the constrained-vs-open-ended structure is not a single-model artifact, but the reliability magnitudes---the paper's central numbers---are characterized on one policy, and the second-policy check is small-$n$ and SFT-only. A broad multi-policy variance study is the natural next step.
\item \textbf{Small $n$.} Held-out results are decision-state emission rates over independent trajectories ($9$--$34$ decision frames from $6$--$9$ trajectories per cell), with $k{=}5$ training seeds; they establish graded existence with resolved confidence intervals, not calibrated end-to-end deployment magnitudes.
\item \textbf{End-to-end $n$ is small.} The end-to-end results (Table~\ref{tab:e2e}) are $5$ rollouts per arm on three cells; they establish the transfer boundary (repair helps when it is the sole blocker) but not calibrated magnitudes. Reaching the warm-served numbers required a load-once serving fix; getting there took us through the same trap the paper warns about---two identical cold-served runs gave challenger $5/5$ then $0/5$ before we removed the per-rollout reload.
\item \textbf{Verifier generality is a capability gradient that tracks scale, not vendor.} Repair results use a Claude verifier. On a 16-frame probe we compared four verifiers (Table~\ref{tab:verifiers}); all agree with Claude $\geq 14/16$ and all make \emph{zero} false-positives ($0/14$: none prematurely calls an incomplete task done, the safe direction for an inference gate). The informative split is the two ``order-placed'' frames that require true done-detection: Claude $2/2$, GPT-4o $1/2$, open Qwen2.5-VL-72B $1/2$, open Qwen2.5-VL-7B $0/2$. Two things follow: done-detection improves with \emph{scale} (open $0/2 {\to} 1/2$ from 7B to 72B), and at comparable scale \emph{open matches closed} (Qwen-72B ties GPT-4o), so the effect is verifier capability, not the open/closed distinction. Claude's $2/2$ edge is a single frame at $n{=}2$---within noise. A calibrated verifier benchmark (hundreds of labeled completion frames) and a second \emph{policy} are the natural next tests.

\begin{table}[h]
\centering
\small
\begin{tabular}{@{}lccc@{}}
\toprule
\textbf{Verifier} & \textbf{Agree w/ Claude} & \textbf{Done-detection} & \textbf{False-pos} \\
\midrule
Claude (closed) & --- & $2/2$ & $0/14$ \\
GPT-4o (closed) & $15/16$ & $1/2$ & $0/14$ \\
Qwen2.5-VL-72B (open) & $15/16$ & $1/2$ & $0/14$ \\
Qwen2.5-VL-7B (open) & $14/16$ & $0/2$ & $0/14$ \\
\bottomrule
\end{tabular}
\caption{Verifier generality on a 16-frame probe (2 completion-positive, 14 negative). Done-detection tracks scale (open 7B$\to$72B: $0/2\to1/2$) and open-72B matches closed GPT-4o; all verifiers are conservative ($0$ false-positives). $n{=}2$ positives---directional, not calibrated.}
\label{tab:verifiers}
\end{table}
\item \textbf{Retracted claim.} We explicitly retract the ``spatial grounding cannot be bought'' boundary of an earlier draft; the replicated held-out value is $0.53{\pm}0.35$ (training-frame $0.36{\pm}0.10$), non-zero in the mean but the least stable across seeds (one of five at $0.00$).
\item \textbf{Verifier-authored labels and frame selection} share one source (frame-selection, teacher, and metric are all the verifier's judgment); we confront this directly where the metric is defined (\S\ref{sec:setup}) and rest any broader claim on the oracle-graded transfer and the multi-verifier agreement rather than the frame-level proxy. Appropriate for a mechanism study, not a data-collection one.
\end{itemize}

\section{Conclusion}
The main lesson is methodological. Agentic computer-use RL numbers are dominated by variance upstream of evaluation---in which data one trains on more than in the seed---and on the hardest cells that variance is bimodal, so its mean and spread mislead. On a verifier-guided repair testbed we make this concrete, twice overturning our own single-run claims. The same discipline underwrites the two findings that survive it: repairability is graded by how constrained the corrective action is, and the frame-level repair transfers end-to-end only when that action is the last thing standing between the policy and completion. For practice: to stabilize an agentic-RL number, control the data draw first and the seed second, average the rest over $k$ runs---and never trust a single run.

\end{document}